\documentclass[aps,prb,twocolumn,groupedaddress,showpacs]{revtex4}

\usepackage{graphics}
\usepackage{amssymb}

\begin{document}


\title{On the corresponding states law of the Yukawa fluid}

\author{Pedro Orea and Yurko Duda
}

\affiliation{\small Programa de Ingenier\'{\i}a Molecular,
Instituto Mexicano del Petr\'{o}leo, Eje Central 152, 07730
M\'{e}xico D.F., M\'exico,}


\begin{abstract}

\noindent

We have analyzed the currently available simulation results, as
well as performed some additional Monte Carlo simulation for the
hard-core attractive Yukawa fluid in order to study its
corresponding state behavior. We show that the values of reduced
surface tension map onto the master curve, and a universal
equation of state can be obtained in the wide range of the
attractive Yukawa tail length after a certain re-scaling of the
number density. Some comparisons with other nonconformal
potentials are presented and discussed.


\end{abstract}

\maketitle

\section {Introduction}

 The investigation of colloid agglomeration,
and phase stability has been a subject of long-standing
theoretical and practical interest
\cite{zaca1,rose1,chav1,cass1,tava1}. It is well accepted that the
range of the effective interaction between particles in different
solvents determines the shape and location of the phase boundaries
of colloidal dispersions \cite{cass1,tava1}. From a theoretical
point of view, the phase transition in colloid-solvent solution is
analogous to that of the single component gas-liquid system. Thus,
the most general approach is to assume a form for the
interparticle effective pair potential. One of the most common
selection is the Yukawa potential,

\begin{equation}
\label{pot}
u(r)=\left\{\begin{array}{ll}\infty, & \mbox{ if $r<\sigma,$}\\
             -\epsilon \sigma exp[-\kappa (r-\sigma)]/r, & \mbox{ if $ \sigma\leq
             r,
             $}
             \end{array} \right.
\end{equation}

\noindent where parameters $\epsilon$ and $\kappa$ define the
depth and range of the potential, respectively; $\sigma$ is a
particle diameter, which is usually taken as a unit length,
$\sigma=1$ .

The range of attraction in a pair potential cannot be changed for
ordinary fluids, but it can be adjusted in colloid-solvent systems
by adding, for instance, a nonadsorbing polymers, ions, or other
solutes \cite{zaca1,cass1,rose1}. Therefore, upon varying the
range of the interaction in the Yukawa potential (\ref{pot}), one
can repoduce the behavior of some real systems.

The phase diagram and interfacial properties of the hard-core
attractive Yukawa (HAY) fluid have been investigated by means of
the theoretical tools of the liquid-state statistical mechanics
\cite{tava1,foff1,blum1,yuka1,wu1,hase1}, and computer simulations
\cite{lomb1,dijk1,duda1,lop1,cail1}. The simulation for measuring
any thermophysical property or phase equilibria for HAY fluid is,
in most cases, time consuming and expensive due to long-range
character of the potential. That is why HAY fluid with long
attraction tails have been recently studied by applying an Ewald
sum \cite{cail1,ewa2}.

On the other hand, it is well-known that in some cases the
application of the corresponding states (CS) law is a very useful
approach to avoid extensive computer calculations
 as well as experimental measurings
\cite{gugg1,foff3,alfe1,cast1,okum1,duni1,noro1}. According to the
CS law the substances have unified equation of state
\begin{equation}
\label{var} P_R = F ({T_R},{\rho_R}),
\end{equation}
in reduced variables of pressure, $P_R=P/P_c$, temperature,
$T_R=T/T_c$, and number density, $\rho_R=\rho/\rho_c$; the
subscript $c$ denotes critical values, and $T \equiv k_B
T/\epsilon$, as usual. In Eq. \ref{var} $F$ is a universal, but
complicated, function. However, it is well-known that the CS
approach is valid only for conformal pair potentials, and Yukawa
potential is not one of them, because its range of attraction
varies independently of the hard-core radius $\sigma$. Taking this
into account, some researchers have applied the so-called extended
CS law for a nonconformal potential, i.e. when Eq. \ref{var}
involves a third parameter \cite{cast1}. Besides, there have been
speculations based on experimental observations and computer
simulations for various nonconformal model potentials that the
gas-liquid coexistence curves may follow an extended CS theory, if
they are plotted in terms of the reduced second virial
coefficient, $B_2^* = B_2(T_c)/B_2^{HS}$ (where $B_2^{HS}$ is the
second virial coefficient of hard sphere fluid), and the reduced
density $\rho_R$ \cite{noro1}. Such considerations were based on
the assumption that second virial coefficient remains practically
constant at the critical point.

However, more recently it has been shown, that $B_2^*$ varies with
the range of attractions, in contrast to the common belief that it
remains practically constant for different pair potentials
\cite{wu1,ped2}. The results of $B_2^*$ for the HAY fluid from
$\kappa = 7$ to $\kappa = 0.5$ are presented in Table 1. Indeed,
its values are far from being constant.

\section {CS law application analysis}

In order to investigate the applicability of the CS law in the
case of HAY fluid, we have analyzed the available simulation data
\cite{lomb1,duda1,cail1,lop1} and performed some additional Monte
Carlo (MC) simulation for relatively long attraction tail, $\kappa
= 1.8$ and $\kappa = 1.5$. Our simulation study has been performed
in the canonical ensemble as described in Refs.
\cite{lomb1,duda1,lop1}. For the phase diagram definition and
surface tension calculations, MC simulations were performed on a
parallelepiped cell with dimensions $L_x=L_y=12$, and $L_z \ge
45$. The pressure of the supercritical HAY fluid was calculated in
a cubic simulation cell, $L_x = 12$. In each simulation, the value
of cut-off radius $r_{cut}$ of the potential (\ref{pot}) was
selected to be $5$. Such value of $r_{cut}$ as well as the system
size have been proved to be sufficient for obtaining accurate
results \cite{duda1,lop1}. Our new results together with the
previously published results, are given in Tables 1 and 2.

The critical parameters for the HAY fluid were calculated by using
the rectiliniear diameter law and the universal value of critical
exponent $\beta = 0.325$. The critical pressures were estimated on
the base of Clausius-Clapeyron equation \cite{okum1}, and together
with other critical parameters are given in Table 3.

In fig.1 we present the dependence of the critical HAY fluid
density $\rho_c$ and pressure $P_c$, on the inverse critical
temperature $T_c^{-1}$ and critical temperature $T_c$,
respectively. Besides our results for the attractive range varying
from $\kappa = 7$ to $\kappa = 1.5$, in the left panel of the
figure we have used the critical data for $\kappa = 0.5$, and
$1.0$ estimated from the recent work of Caillol {\it{et al.}}
\cite{cail1}. All the presented data indicate a clear linear
dependencies described by the following equations:
\begin{equation}
\label{plin} P_c = 0.0228 + 0.0742 T_c,
\end{equation}
\begin{equation}
\label{rholin} \rho_c = 0.2534 + 0.071 \frac{1}{T_c}.
\end{equation}

  Such liner behavior leads to the
constant value of the critical compressibility factor, $Z_c =
 P_c/(\rho_c T_c) = 0.3 \pm 0.01$ in the range of $\kappa$
considered (see also Table 3). This value of $Z_c$ coincides
surprisingly well with its universal value reported in the
literature for different real substances \cite{gugg1}. Actually,
in some cases, this value is being used to estimate the critical
pressure \cite{duni1}.

 In the left part of the Fig.1 we present also the results of the
so-called self-consistent Ornstein-Zernike approximation (SCOZA)
reported in Ref. \cite{foff1,cail1}. As seen, there is a linear
dependence of $\rho_c$ vs $1/T_c$, too. However, the prediction of
the theory deteriorates at the short attractive Yukawa-tails,
$\kappa \ge 4$. This tendency has been discussed in Ref.\cite
{foff1}, where the authors have addressed some tentative reasons
of the SCOZA inaccuracy when the range of the pair potential
narrows. Namely, it was supposed that modification of the closure
condition equation may be needed.

We have also performed preliminary analysis of the available
critical data for the fluids with (i) Sutherland-like attractive
pair interactions, which vary like $r^{-3-t}$ (with $t=3,1$, and
$0.1$ \cite{camp1}), and (ii) Mie $n-6$ $(7<n<32)$ potential
\cite{okum1}. In both cases, there is also a linear dependence
between $\rho_c$ and $1/T_c$, although it is slightly different
from the relation (\ref{rholin}). For the case of the Mie fluid, a
linear function $P_c = P_c(T_c)$ is presented in the right side of
Fig.1. It is interesting to note, that there is a region where
critical parameters of HAY and Mie fluids almost coincide.

On the other hand, the widely used square-well pair potential
shows, according to SCOZA predictions \cite{scho1}, completely
nonlinear behavior of the $\rho_c = \rho_c(T_c^{-1})$, as can be
seen in Fig.1.

If $\rho^L$ and $\rho^G$ denote the liquid and the vapor in mutual
equilibrium at the temperature $T$, respectively, then according
to CS law one should expect $\rho_R^{L,G}$ to be universal
functions of $T_R$. In Fig.2 the liquid-vapor coexistence curves
scaled by the critical values of density and temperature are
presented for different $\kappa$'s. It is seen that most of the
points lie on or near a single curve. More interestingly, the
rectiliniar diameter line, which is defined as \cite{gugg1},

\begin{equation}
\label{recti} \frac{\rho^L+\rho^G}{2 \rho_c} = 1 +
A (1-\frac{T}{T_c}),
\end{equation}

and formed by the simulation results, is very close to the argon
diameter line, with $A=3/4$, depicted as dashed line in Fig. 2.

Since the surface tension, $\gamma$ does not have a reduced
analogue like $T_R$ or $P_R$,  because $\gamma_R =
\gamma/\gamma_c$ is not defined ($\gamma_c=0$), the CS law has
been successfully applied for some substances \cite{gugg1,cast1}
assuming the universality of $\gamma_r = \gamma /(\rho_c^{2/3}
T_c)$ with respect to $T_R$. As we show in Fig.3, all the surface
tension simulation data for the HAY fluid with $\kappa = 7, 5,
1.8$ and $1.5$ map onto the same master curve if presented in
units of $\gamma_r$. Such CS behavior of the reduced surface
tension may answer the questions about the precision of some
recent approximate density functional theories
\cite{weis1,wu2,gloo1}, which predict unexplained nonlinar trends
of $\gamma_r=\gamma_r(T_R)$ function.

The surface tension calculations have been performed through the
pressure tensor \cite{duda1}, therefore, it was natural to suppose
that $\rho_r = \rho/\rho_c^{2/3}$ (instead of $\rho_R$) should be
used in the CS law application for the calculation of the reduced
pressure, $P_R$. We have verified such hypothesis and the results
are presented in Fig.4 and Table 2. As seen, there is a perfect
coincidence of $P_R$ for HAY fluid with $\kappa = 1.8$ and $7$, at
the three reduced temperatures considered.

\section {Conclusions}

 This work presents the detailed analysis of
 the relations between the critical  parameters of HAY fluid with
 variable range of attraction. Such analysis is based on the recently
 reported accurate simulation \cite{duda1,lomb1,cail1,lop1}
 and integral equation data \cite{cail1,foff1}.
 We show that there is a linear relationship between critical pressure
 and critical temperature, as well as, critical density and inverse critical
 temperature for HAY fluid for various interaction ranges.  We have
 also detected such linearity for the bulk critical parameters
 of n-6  \cite{okum1} and Sutherland-like \cite{camp1}  potentials with
 varying attractive tail length.  We have shown that  reduced pressure and
 surface tension of HAY fluid obeys the corresponding state
 law in the range $7<\kappa<1.5$ if the re-scaling critical number
 fluid density  $\rho_c$ is replaced by  $\rho_c^{2/3}$. Finally,
 our findings permit us to conclude that phase diagram,
 pressure and surface tensions
 of the HAY fluid can be successfully obtained applying
 the properly modified corresponding state theory.
 Thus, we introduce a new criteria for the estimation and
 improving of the theoretical approaches
 \cite{tava1,foff1,yuka1,wu1,weis1,blum1,wu2,gloo1}, i.e. now the
 predictive accuracy of different theories can be verified by applying
 of CS law. It is very probable, that other fluids  interacting
 via an attractive nonconformal potential \cite{okum1,duni1,camp1} can be
 studied with the same CS approach. To verify  such hypothesis additional
 accurate simulations are desirable.


\section{Acknowledgments}

We gratefully acknowledge the financial support of the Instituto
Mexicano del Petr\'oleo under the Projects $D.31519/D.00406$.


\newpage

\begin{table}
\caption{Canonical Monte Carlo data for the phase equilibrium and
surface tension of the HAY fluid with $\kappa = 1.5$ and $1.8$.}

\label{table1}
\begin{tabular}{cccccc}
\hline\hline
\hspace{1.0cm} $T$\hspace{1.0cm}  & $\rho_{L}$\hspace{0.7cm} & $\rho_{G}$\hspace{0.7cm} & $\gamma$\hspace{0.1cm}  \\

\hline \vspace{-0.3cm}
\\ &   $\kappa = 1.5$ & & \\   \hline
      1.10   & 0.730    & 0.0207           & $0.760_{20}$  \\
       1.15   & 0.702    & 0.0290           & $0.633_{15}$  \\
       1.20   & 0.665    & 0.0415           & $0.515_{20}$  \\
       1.25   & 0.633    & 0.0525           & $0.390_{15}$  \\
       1.30   & 0.590    & 0.0725           & $0.271_{10}$  \\
       1.35   & 0.545    & 0.0930           & $0.165_{15}$  \\ \hline\vspace{-0.3cm}

 \\ &  $\kappa = 1.8$ & & \\   \hline

   0.90  & 0.754    & 0.0220            & $0.612_{15}$  \\
   0.95  & 0.715    & 0.0313            & $0.475_{15}$  \\
   1.00  & 0.672    & 0.0463            & $0.360_{10}$  \\
   1.05  & 0.624    & 0.0650            & $0.241_{10}$  \\
   1.10  & 0.565    & 0.0880            & $0.140_{10}$  \\
   1.13  & 0.530    & 0.1060            & $0.088_{10}$  \\
   1.15  & 0.500    & 0.1432            & $0.050_{10}$  \\ \hline

\end{tabular}
\end{table}

\newpage

\begin{table}
\caption{ Canonical Monte Carlo results for the supercritical
pressure, $P$,  of the HAY fluid with $\kappa = 7$ and $1.8$.}

\label{table1}

\begin{tabular}{c|ccc|ccc}
\hline\hline
\hspace{1.0cm}& &$\kappa=7.0$\hspace{1.0cm}&&\hspace{1.0cm}&$\kappa=1.8$\\
\hline
$\hspace{1.0cm}\rho$ \hspace{0.2cm} &$T=0.432$\hspace{1.0cm}& $0.50$\hspace{1.0cm}&$0.828$\hspace{0.5cm}& $T=1.0$\hspace{0.5cm}& $1.425$\hspace{0.5cm}& \hspace{0.3cm}$2.36$ \\
\hline

0.2 & 0.052 & 0.077 & 0.188 & 0.130  & 0.193 & 0.487  \\
0.4 & 0.070 & 0.136 & 0.478 & 0.160  & 0.347 & 1.280  \\
0.5 & 0.077 & -     &    -  & 0.210  & 0.530 & 2.040  \\
0.6 & 0.090 & 0.235 & 0.900 & 0.450  & 0.962 & 3.420  \\
0.7 & 0.128 & 0.351 & 1.400 & 1.000  & 1.855 & 5.590  \\
0.8 & 0.240 & 0.580 & 2.200 & 2.400  & 3.631 & 9.450  \\
0.9 & 0.530 & 1.090 & 3.500 & -      & -     & -      \\

\end{tabular}
\end{table}

\newpage

\newpage

\begin{table}
\caption{Critical parameters of the HCY fluid
 with different values of the range parameter $\kappa$. }

\label{table1}

\begin{tabular}{cc|ccccccc}
\hline\hline $\hspace{0.3cm}\kappa$\hspace{0.2cm} &
Ref.\hspace{0.3cm} &\hspace{0.3cm} $T_c$\hspace{0.3cm}  &
$\hspace{0.3cm}\rho_c$\hspace{0.5cm} & \hspace{0.4cm} $B_2^*$
\hspace{0.5cm}
& $P_c$\hspace{0.5cm}  &  $Z_c$ \hspace{0.5cm}  \\

\hline
 0.5   & \cite{cail1}& 6.958 & 0.260  & -6.474 &  -    &  -       \\
 1.0   & \cite{cail1}& 2.460 & 0.280  & -6.290 &  -    &  -       \\
 1.5   &             & 1.451 & 0.304  & -6.277 &  0.128   &  0.300   \\
 1.8   &             & 1.180  & 0.313  &-6.212&  0.112   &  0.300   \\
 1.8   & \cite{cail1} & 1.189 & 0.317  &-6.122&  -   &  -   \\
 1.8   & \cite{lomb1}& 1.177  & 0.313  &-6.242&  -   &  -   \\
 2.0   & \cite{lop1} & 1.050 & 0.322  &-6.232&  0.102   &  0.305   \\
 2.5   &  \cite{lop1}& 0.840 & 0.336  & -6.206&  0.086   &  0.303   \\
 3.0   & \cite{lop1} & 0.721 & 0.356  & -6.028&  0.076   &  0.296   \\
 4.0   & \cite{duda1}& 0.581 & 0.380  &-5.795&  0.066   &  0.302   \\
 5.0   & \cite{duda1}& 0.500 & 0.393  &-5.684&  0.060   &  0.305   \\
 6.0   & \cite{duda1}& 0.448 & 0.412  &-5.590&  0.055   &  0.298   \\
 7.0   & \cite{duda1}& 0.414 &  0.422 & -5.402&  0.053   &  0.302
 \\\hline .

\end{tabular}
\end{table}

\newpage

.\vspace{19cm}

\section*{Figure Captions}

\noindent {\bf Fig.1} Critical density $\rho_c$ as a function of
inverse critical temperature $T_c^{-1}$ (left panel); critical
pressure $P_c$ as a function of critical temperature $T_c$ (right
panel). The results of the HAY fluids are presented by fill
circles (MC results), and dotted line (SCOZA \cite{cail1,foff1}).
The squares depict Molecular Dynamic simulation data of Ref.
\cite{cail1} for $n-6$ potential; dashed line represents SCOZA
theory predictions for the square-well fluid \cite{scho1}. \\

\noindent {\bf Fig.2} Reduced vapor-liquid coexistence curves for
HAY fluid at different ranges of attractions: $\kappa=7$, and $5$
(Ref. \cite{duda1}); $\kappa=1.8$ and $1.5$ (present work); $\kappa=1$ (Ref. \cite{cail1}).   \\

\noindent {\bf Fig.3} Reduced surface tension $\gamma_r$ as a
function of reduced temperature $T_R$ at different ranges of
attraction. The meaning of the symbols is the same as in Fig.2.  \\

\noindent {\bf Fig.4} Reduced pressure $P_R$ as a function of
reduced HAY fluid density $\rho_r = \rho/\rho_c^{2/3}$ at three
reduced temperatures, $T_R = 1.042$, 1.208, and $2.0$.
Two ranges of attraction, $\kappa = 7$ (squares) and $1.8$ (circles) are
considered.\\


\newpage

\centering

\resizebox{0.468\textwidth}{0.27\textheight}{\includegraphics{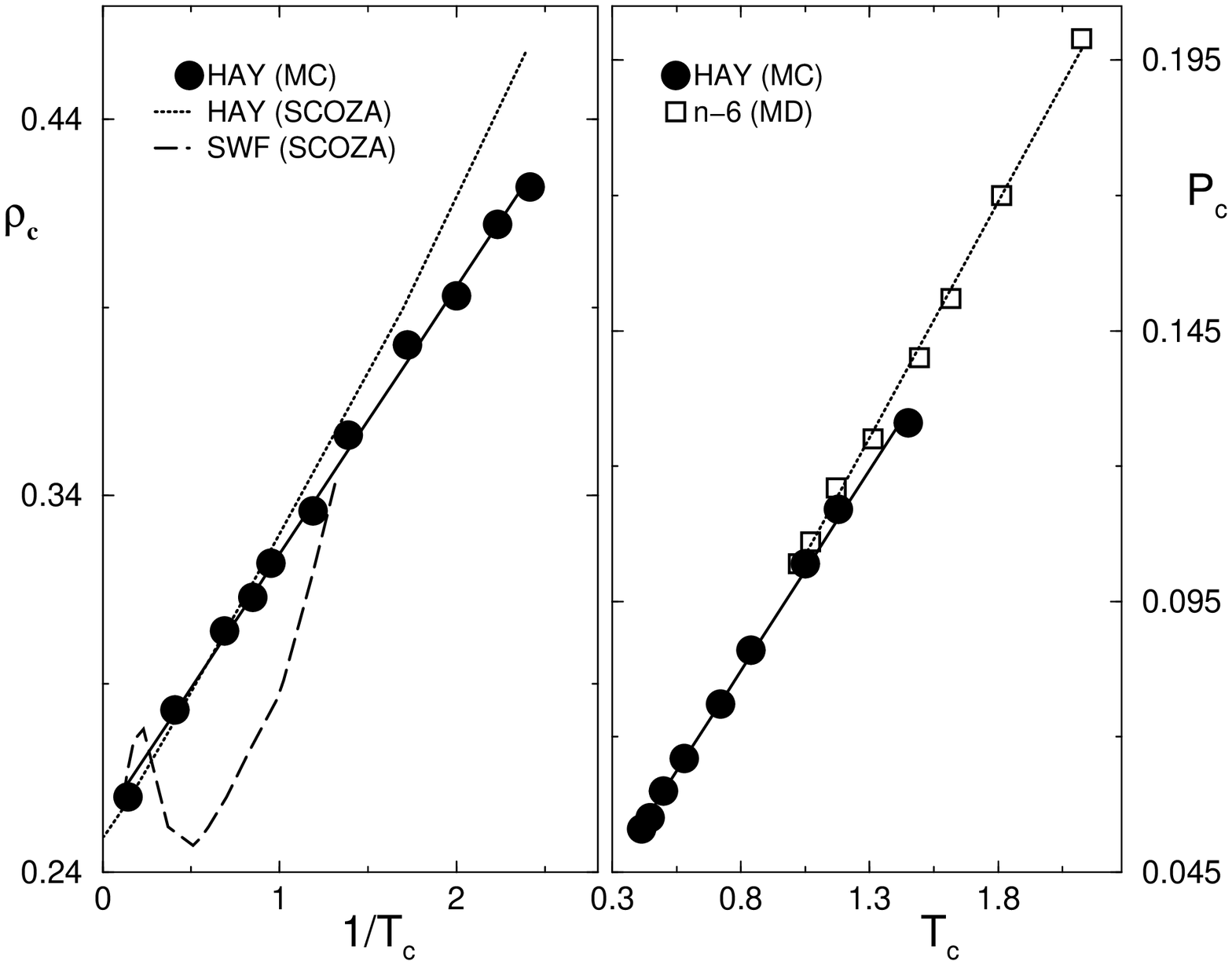}}
\vspace{0.5cm}

\centerline{Fig. 1}

\vspace{0.5cm}

\newpage

\centering
\resizebox{0.39\textwidth}{0.255\textheight}{\includegraphics{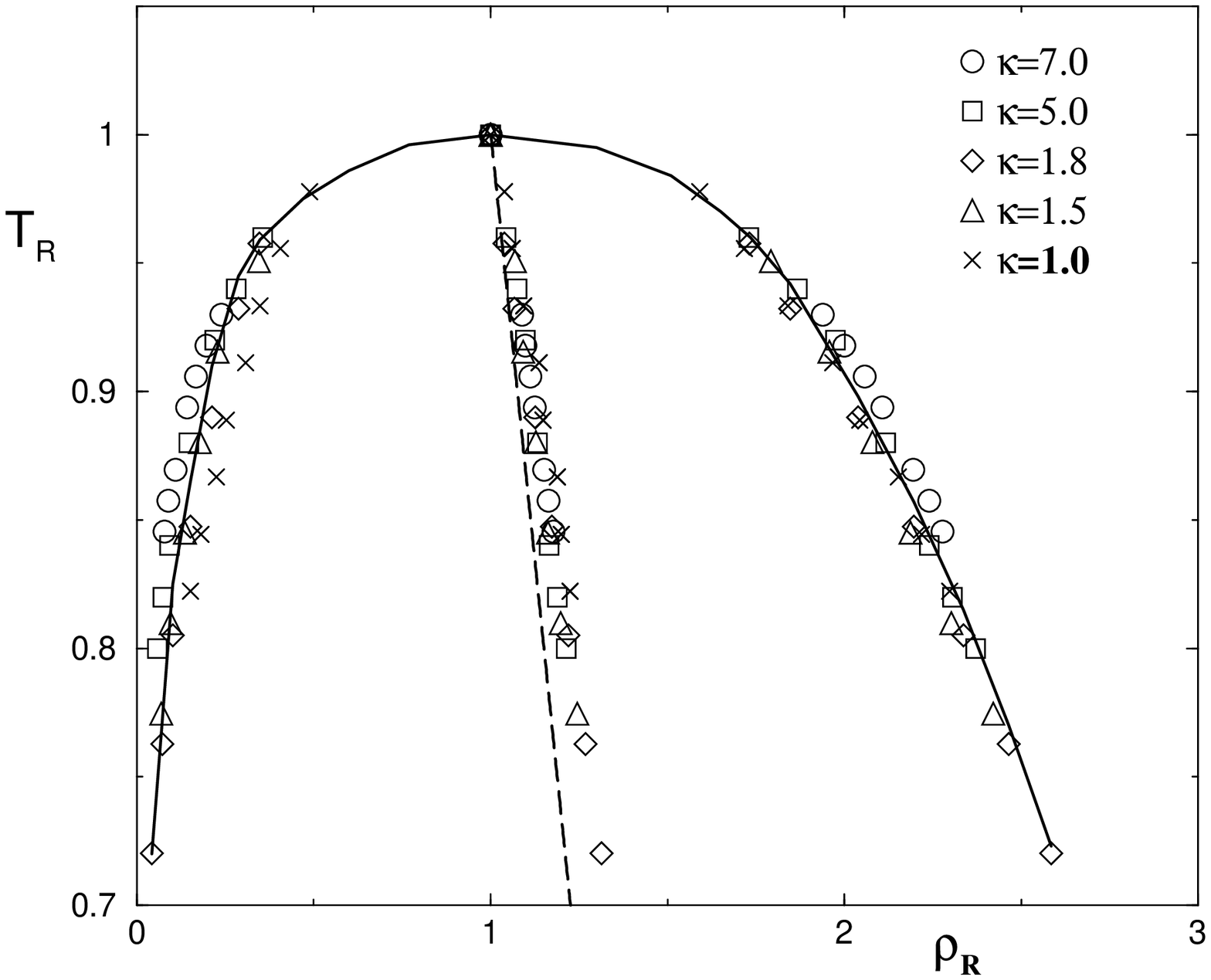}}

\vspace{0.5cm}

\centerline{Fig. 2}

\newpage

\centering
\resizebox{0.39\textwidth}{0.255\textheight}{\includegraphics{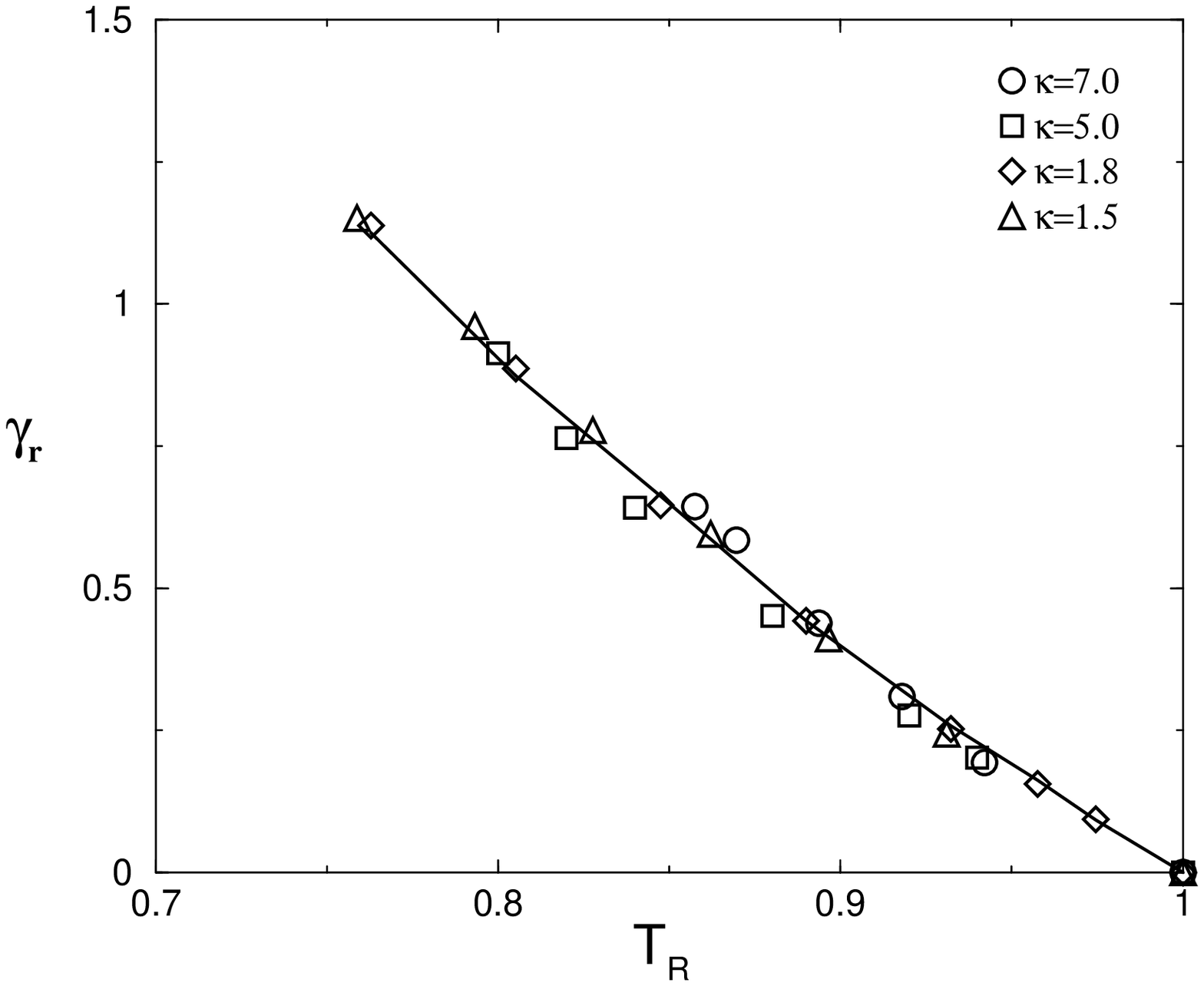}}

\vspace{.5cm}

\centerline{Fig. 3}\vspace{.5cm}

\newpage

\centering
\resizebox{0.39\textwidth}{0.255\textheight}{\includegraphics{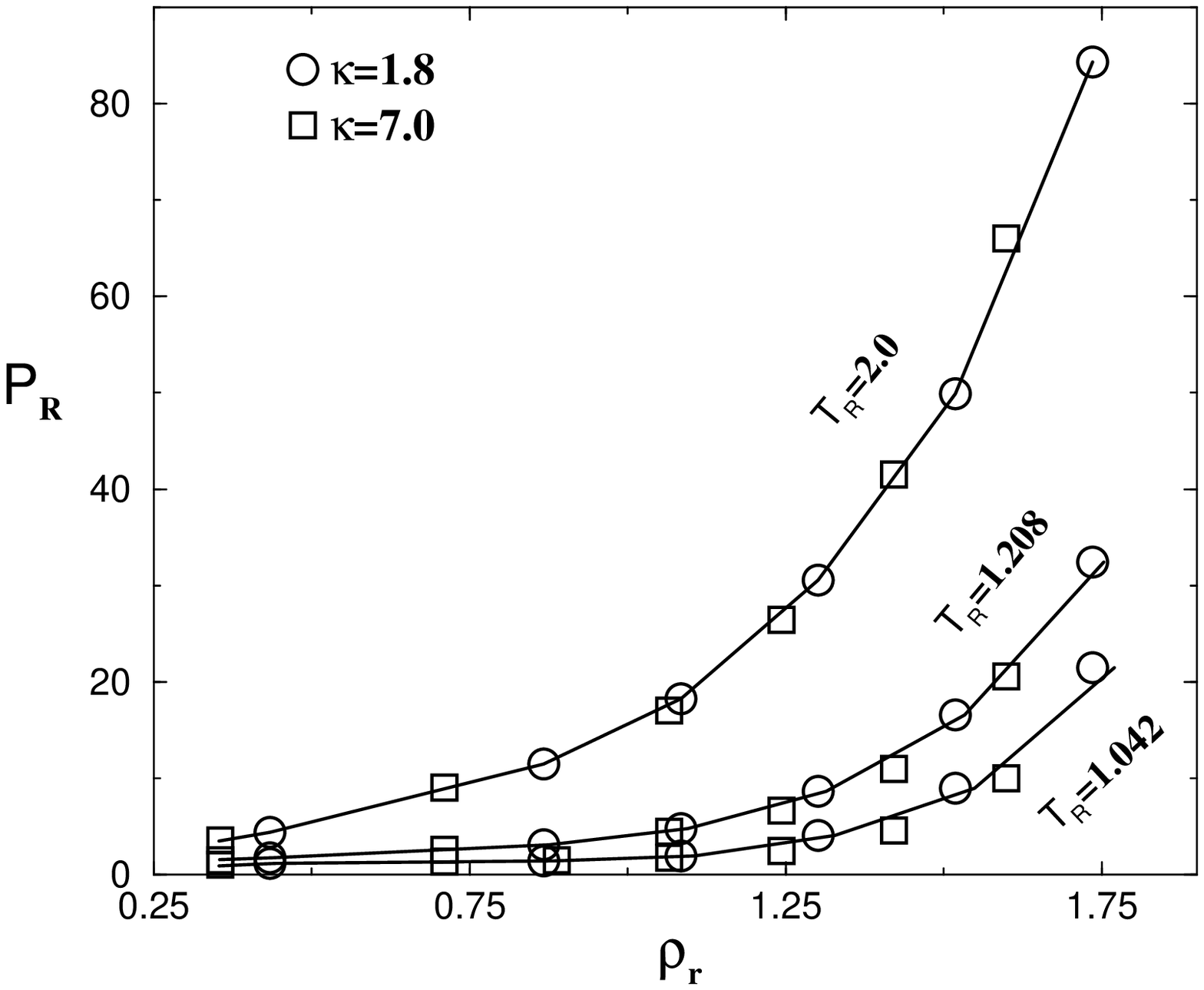}}

\vspace{0.5cm} \centerline{Fig. 4}



\end{document}